\documentclass[a4paper,aps, prx, onecolumn, nofootinbib, 11pt, superscriptaddress, longbibliography, tightenlines]{revtex4-2}

\usepackage[utf8]{inputenc}
\usepackage[english]{babel}
\usepackage[a4paper, margin=2.25cm]{geometry}
\usepackage{graphicx}
\usepackage{dsfont}
\usepackage{amsmath, amssymb}
\usepackage{amsfonts}
\usepackage{physics}
\usepackage{bm}
\usepackage[dvipsnames]{xcolor}
\usepackage{tikz}
\usetikzlibrary{matrix, arrows.meta,positioning}
\usepackage[colorlinks=true,linkcolor=teal,citecolor=Maroon,urlcolor=Maroon]{hyperref}
\usepackage{tabularx}
\usepackage{tcolorbox}
\usepackage[normalem]{ulem}
\usepackage{needspace}

\usepackage{booktabs, amsthm}
\usepackage{stmaryrd}
\usepackage{makecell}
\usepackage{cleveref}
\crefformat{equation}{(#2#1#3)} 
\usepackage[noend,linesnumbered, ruled, vlined]{algorithm2e}

\newcommand{\secref}[1]{Section~\hyperref[#1]{\ref*{#1}}}
\newcommand{\appref}[1]{Appendix~\hyperref[#1]{\ref*{#1}}}
\newcommand{\tabref}[1]{Table~\hyperref[#1]{\ref*{#1}}}
\newcommand{\figref}[1]{fig.~\hyperref[#1]{\ref*{#1}}}
\newcommand{\sfigref}[2]{fig.~\hyperref[#1]{\ref*{#1}(#2)}}

\newtheorem{theorem}{Theorem}

\theoremstyle{remark}
\newtheorem{remark}{Remark}

\newcommand{\subsubhead}[1]{\needspace{6pt}
\par
\vspace{0.5\baselineskip}\noindent\textbf{#1}}

\begin{document}

\title{Efficiently tracking errors and outcomes in universal quantum circuits}
\author{Thomas R. Scruby}
\email{t.r.scruby@gmail.com}
\affiliation{\footnotesize Okinawa Institute of Science and Technology, 1919-1 Tancha, Onna, Okinawa, Japan}
\affiliation{\footnotesize Iceberg Quantum}
\author{Julio C. Magdalena de la Fuente}
\affiliation{\footnotesize Dahlem Center for Complex Quantum Systems, Freie Universit\"at Berlin, 14195 Berlin, Germany}
\affiliation{\footnotesize School of Physics, The University of Sydney, NSW 2006, Australia}

\begin{abstract}
\vspace{6pt}

\begin{center}
    \textbf{Abstract}
\end{center}

We present a method for efficient simulation of error propagation and decoding in universal fault-tolerant quantum circuits under stochastic Pauli noise, requiring only that the circuit has a high fault distance and that all logical non-Clifford operations are implemented by physical T, CS and CCZ gates.
This is achieved using modified Pauli frame-based error tracking in which the frame contains no logical information.
Decoding success is identified by verifying that the Pauli frame is still logically trivial after decoding.
Using the results of Ref.~\cite{companion-nonclifford} and the observation that the distribution of non-logical measurement outcomes in the circuit is independent of the logical state, we show that we can accurately sample all detector outcomes. Our approach can also simulate circuits containing adaptive operations conditioned on logical measurements, provided these measurements have a known outcome distribution. This applies to the majority of adaptive operations such as state teleportation and preparation by measurement. 
Our method only detects decoding failures and so is consistent with complexity-theoretic expectations as it cannot be used to sample from the noiseless outcome distribution of the logical circuit.
\end{abstract}

\maketitle

\section{Introduction}
\label{section:introduction}
It is widely accepted that quantum error-correcting codes (QECCs) will be necessary components of any meaningful implementation of a complex quantum algorithm or deep quantum circuit~\cite{nielsen_quantum_2011}.
The question of which codes are most practically useful is a challenging one, and while code parameters or implementable logical operations can often be studied analytically~\cite{breuckmann_ldpc_2021}, the actual performance of a code given a particular circuit implementation, noise model and decoder must be assessed via simulation. 

The recent proliferation of publicly available simulation tools and decoding algorithms~\cite{gidney_stim_2021,roffe_decoding_2020,higgott2021pymatchingpythonpackagedecoding} has made answering these questions much easier in certain cases, but in other cases limitations still exist. Most notably, \texttt{stim} and other tools that work via direct circuit simulation have traditionally been limited to the case of physical Clifford circuits, where this simulation can be performed efficiently~\cite{Aaronson2004circuits}. However, precisely because of this simulability, any practically interesting quantum circuit must necessarily contain non-Clifford operations and without alternative methods for investigating logical failure rates the robustness of these circuits cannot be tested. 
A large amount of work has recently been dedicated to addressing this problem, and the simulation methods that have been proposed can broadly be divided into two categories. Either they provide an efficient method applicable only to a non-universal class of circuits~\cite{surti2025efficientsimulationlogicalmagic, yugopaper, cercelescu2026polynomialtimesimulationnoncliffordquantum, wang2026magnasimulatingmagicstate, xiao2026efficientfidelitysimulationhighrate}, or they provide an optimised but still inefficient method applicable to universal circuits~\cite{Sutcliffe_2024, haenel2026tsimfastuniversalsimulator,umbrarescu2026syqmamemoryefficientsymbolicexact,li2025softhighperformancesimulatoruniversal}.

In this work we present, to our knowledge, the first simulation technique that is both efficient and applicable to universal quantum circuits. 
This is not in conflict with complexity-theoretic expectations as we accurately sample only from the difference between the noisy and noiseless circuit outputs, and not from the true output statistics in the noiseless case. 
In this way we can study the failure rate of the circuit without obtaining any knowledge of the idealised output. 
The method is applicable to any fault-tolerant quantum circuit subject to stochastic Pauli noise where
\begin{itemize}
    \item the spacetime decoding problem has a high distance and
    \item all logical non-Clifford operations are implemented by \textit{third-order blocks}, which are circuits composed of $X$ and $Z$ initialisations and measurements, as well as CX, T, CS, and CCZ gates.
\end{itemize}
The first restriction is fundamental to our approach, but we expect the second restriction can be lifted with further work. 
Our approach does not require tracking of Clifford operators and instead functions by modifying standard Pauli-frame tracking methods and supplementing them with efficient methods for propagation through non-Clifford operations, described in detail in our previous work~\cite{companion-nonclifford}.

In the rest of this paper we describe our simulation technique, first at a high level in \cref{section:overview}, then in more detail in \cref{section:blueprint}. We then discuss routes towards scalable simulation of failure rates in large quantum circuits in \cref{section:discussion}.

\section{Overview of the approach}
\label{section:overview}
To begin, we provide a high-level overview of our proposed simulation strategy. The aim of this section is to provide the reader with a broad conceptual understanding of how the simulator works, and how it can be both accurate and efficient, while minimising technical discussion. Full details are provided in later sections.

\subsection{Setup and Scope}

In a fault-tolerant implementation of a logical circuit the quantum information is encoded in a codespace, $\mathcal{C}\subseteq\mathcal{H}$, where $\mathcal{H}$ denotes the physical Hilbert space formed by all qubits in the quantum computer.
Logical states in $\mathcal{C}$ will be written $\ket{\overline{\psi}}_\mathcal{C}$.

The physical circuits are decomposed into \textit{logical blocks}~\cite{psilogicalblocks}, which implement maps from one encoding $\mathcal{C}$ to another $\mathcal{C}'$,
\begin{equation}
    \mathcal{M}: \ket{\overline{\psi}}_\mathcal{C} \rightarrow \ket{\overline{\psi'}}_\mathcal{C'}.
\end{equation}
These blocks can include adaptive physical or logical operations conditioned on measurement outcomes. 
In this work, we consider circuits composed of logical blocks satisfying some additional properties:
\begin{itemize}
    \item \textit{well-defined logical action:} In the absence of errors, each block implements a well-defined logical gate.
    We consider circuits composed of
    \begin{itemize}
        \item Clifford blocks: Used to implement logical unitary Clifford gates or logical Pauli measurements
        \item Third-order blocks: Used to implement logical non-Clifford unitaries or logical Clifford measurements.
    \end{itemize}
    By ``third-order block'', we mean a fault-tolerant logical operation composed of Pauli $X$/$Z$ initialisations and measurements, and CX, T, CS and CCZ gates. These are common building blocks of proposals for  practical fault-tolerant architectures for universal quantum computers~\cite{gidney2025factor2048bitrsa, pinnacle, tourdegross, oratomicarchitecture}.
    \item \textit{fault tolerant:} Localised errors have no effect on the logical action. We only consider errors on quantum gates and assume the classical computation and communication to be noiseless.
    \item \textit{efficiently decodable:}    
    There exist sets of measurement outcomes that can be used as input to an efficient algorithm that computes corrections for errors. We say that decoding succeeds if the combined action of the noisy circuit plus correction matches that of the noiseless circuit.
\end{itemize}

In this setting, we would like an efficient method to answer the following question:
\begin{center}
\begin{tcolorbox}[
  colback=white!95!black,
  colframe=white!20!black,
  boxrule=0.5pt,
  width=0.7\linewidth,
  arc=2mm,
  boxsep=0pt,
  halign=center
]
    For a given circuit composed of noisy logical blocks, what is the probability that decoding succeeds?
\end{tcolorbox}
\end{center}

Notice that the knowledge we require in order to answer this question is actually fairly minimal compared to the full knowledge about the system's state at any time within the circuit.
We do not need to know the true action of the circuit in the noiseless case, or how this action differs in the event of a decoding failure.
We only need to know \textit{if} it differs.
Our approach is to achieve that by sampling errors in the circuits and tracking their propagation correctly.
It succeeds as long as
\begin{itemize}
    \item it accurately tracks the propagation of all errors that do not cause decoding failures
    \item it accurately detects the existence of any error that does cause a decoding failure.
\end{itemize}
A high-level reason that this approach works in the fault-tolerant setting is that if the errors act in correctable regions of the underlying error-correcting code their effect must be trivial on the logical information by the Knill-Laflamme conditions~\cite{knill1997theory} and their operator-algebraic generalizations~\cite{kribs2006operatorQEC}.
As a consequence, the effect of the errors can be inferred locally and independently of tracking the logical information.
In this way, our approach explicitly separates the computationally intractable parts of a universal fault-tolerant circuit from the tractable parts.

\subsection{A sketch of the simulator}

The two tasks of the simulator are separated into \emph{error propagation} and \emph{decoding}.

\subsubsection*{Tracking errors}

Within Clifford circuits Pauli errors can be efficiently propagated using the stabilizer formalism~\cite{Aaronson2004circuits}.
Recent works~\cite{yugopaper,fechisin2026cliffordhierarchystabilizerformalismapplications,cercelescu2026polynomialtimesimulationnoncliffordquantum} have generalised this approach to certain classes of non-Clifford circuit, but it is not expected that this method can be efficient in the general case.
In Ref.~\cite{companion-nonclifford} we provide an efficient method to propagate Pauli errors through third-order blocks, allowing us to combine it with standard error-propagation methods for the Clifford blocks.
In the following, we review the structural properties of tracking the errors reliably through a third-order block.

The error-correction conditions~\cite{knill1997theory} ensure that after measuring a complete set of generators of a stabilizer code the state is of the form $P_e\ket{\overline{\psi}}$ for some Pauli error $P_e$.
Hence, the action of any logical block can be written as \begin{equation}
    \mathcal{M}: P_e\ket{\overline{\psi}}_\mathcal{C} \rightarrow P_e'\ket{\overline{\psi'}}_\mathcal{C'}.
\end{equation}
If $P_e$ acts on a subset of qubits that does not support any logical operators, and if this property is preserved throughout the transformation, $P_e \rightarrow P_e'$, then the transformation of the global state \emph{must} have been independent of $\ket{\overline{\psi}}$.
The action of the logical block $\mathcal{M}$ then factorizes,
\begin{equation}
    \label{eq:factors}
    \mathcal{M}(P_e\ket{\overline{\psi}}) = \mathcal{M}_1(P_e)\mathcal{M}_2(\ket{\overline{\psi}}).
\end{equation}
Our simulator only computes $\mathcal{M}_1(P_e)$, conditioned on the guarantee that the logical information could not have been impacted in the process.

In Ref.~\cite{companion-nonclifford} we provide a general method to compute $\mathcal{M}_1$ for fault-tolerant logical blocks that implement logical diagonal gates in the third level of the Clifford hierarchy on CSS codes. We call the physical circuits that implement them third-order circuits.
These blocks are common in many fault-tolerant architectures that can perform universal quantum computation.
A \emph{contractibility condition} (condition C)~\cite{companion-nonclifford} formalizes the cleanability condition above within the circuit and guarantees that the logical channel remains unaffected.
When condition C does not hold we declare failure, and do not need to track the residual errors.
In general, the condition can only be checked approximately and in such cases we can only upper bound the failure rate.

\subsubsection*{Decoding and declaring failure}
In addition to tracking errors correctly in the simulation, we must also be able to properly sample from check measurement outcomes for use in decoding, as well as detect when decoding failures cause logical errors. Both of these things are non-trivial for a general circuit, as it can be hard to sample from the distribution of measurement outcomes. We get around this issue by using a Pauli frame simulator (of the type used in \texttt{stim}~\cite{gidney_stim_2021}).
It does not sample measurement outcomes directly and instead samples flips relative to the outcomes of a noiseless reference circuit. We get around the need for access to such reference outcomes in two different ways. 

For decoding it is important to learn real measurement outcomes since the technique described in \cite{companion-nonclifford} requires a sparse $X$ error configuration on the block's input state. To achieve this one must apply a correction to the state immediately before the block that attempts to set all stabilizer outcomes to $+1$. If we do not know which stabilizer outcomes are $-1$ then this is obviously not possible. We obtain reference outcomes for stabilizers using a priori knowledge about fault-tolerant circuits; for example, on a perfect codestate all stabilizers are guaranteed to have measurement outcome $+1$.

When deciding logical success or failure we cannot hope for a noiseless reference outcome to compare to (access to this would remove the need to run the quantum circuit entirely), but this is also not a problem as we do not care to learn the true outcome, only to learn about flips of this outcome. The Pauli frame simulation approach employed by \texttt{stim} can produce such flips even in noiseless circuits, for example due to successive anticommuting logical measurements, but we observe that a simple modification to this technique allows one to ensure that flips of logical outcomes occur \textit{only} in the event of decoding failure. 

\section{Simulator blueprint}
\label{section:blueprint}
In this section we provide a full description of our proposed simulator, beginning at the level of physical operations, then discussing the treatment of different types of logical blocks, followed by how the circuit can be decoded and logical success or failure can be decided.
At the end we discuss how, in cases of interest, our method continues to be valid for logical circuits that contain gates conditioned on measurement outcomes.

\subsection{Physical operations}
\label{subsection:physical_operations}

Our Pauli tracking method is an extension of the standard Pauli frame tracking approach employed by \texttt{stim}~\cite{gidney_stim_2021}. It builds on the symplectic representation of Pauli operators based on binary vectors.
We define a \emph{Pauli frame} for a set of $n$ qubits as a pair of length-$n$ binary vectors $\bm{x}, \bm{z}\in\mathbb{Z}_2^n$ that represent the products of Pauli operators applied to the qubits throughout the circuit by
\begin{equation}
    X[\bm{x}] := \bigotimes_i X_i^{x_i}, ~~~ Z[\bm{z}] := \bigotimes_i Z_i^{z_i}
\end{equation}
We will sometimes refer to the vectors $\bm{x}$ and $\bm{z}$ as ``operators'', always meaning the operators they describe via this definition. We will write the associated pair of binary variables for a qubit $i$ as $[x_i,z_i]$ and the full Pauli frame as $\mathcal{F} = [\bm{x},\bm{z}]$.
Given a stabilizer code, the Pauli frame decomposes non-uniquely according to $\mathcal{F} = \bm{p}_l + \bm{p}_s + \bm{p}_e$ into \emph{logical}, \emph{stabilizer} and \emph{error} components, each of which is a length-$2n$ binary vector, in general.
For CSS codes, they naturally decompose into length-$n$ $X$ and $Z$ components, respectively.
Throughout, $+$ denotes binary addition.
Gates and measurements in a Pauli frame simulation of a quantum circuit are represented by functions that transform the Pauli frame either deterministically or probabilistically.
We specifically consider circuits composed of the operations listed below.
We first give definitions of how they transform the Pauli frame, then justify these and explain why tracked errors transform correctly in the following sections.

\subsubhead{Pauli errors and corrections $(X,Y,Z)$}
\[
    \left.
    \begin{split}
        &X: [x_i,z_i] \rightarrow [x_i + p,z_i] \\
        &Y: [x_i,z_i] \rightarrow [x_i + p,z_i + p] \\
        &Z: [x_i,z_i] \rightarrow [x_i,z_i+p] 
    \end{split}
    \right\}
    \textrm{possibly random bit}~ p
\]
This simply captures the idea that $X^2 = Y^2 = Z^2 = \mathbb{I}$ and that, up to phases, $XZ = Y$. For a Pauli error, $p=1$ with probability equal to the probability of that Pauli error occurring in our chosen noise model. When simulating an explicit correction $p=1$ deterministically.

\subsubhead{Clifford unitaries $(U_\textrm{C})$}
\[ 
    U_\textrm{C}: [\bm{x},\bm{z}] \rightarrow [\bm{x}',\bm{z}'], ~ \textrm{ for } U_\textrm{C}(\phi X[\bm{x}]Z[\bm{z}])U_\textrm{C}^\dagger = \phi' X[\bm{x}']Z[\bm{z}']
\]
Any $n$-qubit Pauli operator can be uniquely written as $\phi X[\bm{x}]Z[\bm{z}]$ for some $\bm{x}$ and $\bm{z}$ and phase $\phi$, and any $n$-qubit Clifford operator can be defined as an automorphism of the $n$-qubit Pauli group.
Together these give a well-defined action of Clifford operators on the Pauli frame.
The transformation $[\bm{x},\bm{z}] \rightarrow [\bm{x}',\bm{z}']$ is linear and the matrix implementing it is the symplectic representation of $U_\textrm{C}$.
Importantly, we treat physical Pauli operators that are placed in the circuit deterministically as Clifford unitaries.
As a consequence, they do \emph{not} change the frame -- in contrast to corrections.

\subsubhead{Qubit resets $(R_X,R_Z)$ and measurements $(M_X, M_Z)$}\\
We consider only single-qubit resets but allow for multi-qubit measurements.
\[
    \left.
    \begin{array}{c}
        R_{Z_i}: ~~ [x_i,z_i] \rightarrow [0,z_i+p] \\
        R_{X_i}: ~~ [x_i,z_i] \rightarrow [x_i+p,0] \\
        \\
        M_{Z[\bm{z}']}:
        \begin{array}{c}
            [\bm{x},\bm{z}] \rightarrow [\bm{x},\bm{z} + p\bm{z}'] \\
            \textrm{outcome } \rightarrow (-1)^{\bm{x} \cdot \bm{z}'}m
        \end{array} \\
        M_{X[\bm{x}']}:
        \begin{array}{c}
            [\bm{x},\bm{z}] \rightarrow [\bm{x} + p\bm{x}',\bm{z}] \\
            \textrm{outcome } \rightarrow (-1)^{\bm{x}' \cdot \bm{z}}m
        \end{array} \\
    \end{array}
    \right\} 
    \begin{array}{c}
        \textrm{ fully random bit } p \\
        \textrm{ noiseless outcome } m,
    \end{array}
\]
where $\cdot$ denotes the standard inner product of vectors.
The randomisation of the $Z$ (resp. $X$) component of the Pauli frame by $M_{Z[\bm{z}']}$ captures the idea that this is a projective operation that can destroy information about whether or not $Z[\bm{z}']$ was applied to the qubit prior to the operation, or alternatively, that each measurement projects onto an eigenstate of $Z[\bm{z}']$ and so applying $Z[\bm{z}']$ to the state post-measurement should not alter it.
The sign $(-1)^{\bm{x}' \cdot \bm{z}}$ or $(-1)^{\bm{x} \cdot \bm{z}'}$ captures whether the measurement outcome in our simulation is flipped relative to some noiseless reference measurement $m$.
Given knowledge of $m$ we can compute this noisy outcome, but even in the case where we do not have access to such a reference measurement the Pauli frame still contains information about this flip. 

\subsubhead{Third-level diagonal non-Clifford gates}\\ 
In the most generally applicable version of our simulator we do not consider the local effect of individual non-Clifford gates on physical qubits, but rather the effect of a non-Clifford logical block on the entire code for which we provide a general method in~\cite{companion-nonclifford}.
This method is based on the observation that the distribution of combined measurement outcomes and errors in fault-tolerant third-order circuits is subject to affine linear constraints that can be computed efficiently.
Within the subspace every consistent error configuration happens with the same probability.
This allows us to model the circuit by sampling a set of random \emph{twisted errors} from a flat distribution over a subspace of errors.
This includes an offset error that has to be computed for each $X$-error configuration and added into the circuit to faithfully model the measurement distribution of a third-order circuit.

In App.~\ref{app:ccz_sampling} we present an alternate method to sample twisted errors that uses only a set of gate-local update rules for the Pauli frame.
We prove that for inter-block CCZ circuits, e.g. transversal or copy-cup CCZ gates, this method correctly samples from the full twisted error distribution, allowing for simulation of non-Clifford blocks in these circuits with the same overhead as for Clifford circuits with \texttt{stim}.
In general, however, the offset error is not captured reliably in this way and a more global method must be used, as explained in Ref.~\cite{companion-nonclifford}.

\subsection{Logical blocks}

We now consider all operations of interest at the logical level and in each case ask how the frame transforms and which measurement outcomes can be learned accurately.

\subsubsection*{Pauli eigenstate preparation}

For ease of discussion we focus specifically on the case of preparing $\ket{\overline{0}}^{\otimes k}$, but other cases -- such as preparing $\ket{\overline{+}}^{\otimes k}$, or code switching operations that prepare new logical qubits in a Pauli basis -- function in exactly the same way.

\subsubhead{What happens to the frame?}\\
Consider the standard method of preparing the logical $\ket{\overline{0}}^{\otimes k}$ state of an $[\![n,k,d]\!]$ CSS code, where we first prepare all $n$ physical qubits in $\ket{0}$ then measure all $X$ stabilizers. In the noiseless case, and using the definitions from the previous subsection, resetting $n$ qubits to $\ket{0}^{\otimes n}$ gives a Pauli frame $\mathcal{F} = [\bm{0}, \bm{z}]$ where $\bm{0}$ and $\bm{z}$ are the length-$n$ all-zeros vector and a length-$n$ random binary vector respectively. Each $X$ stabilizer we measure then gets multiplied into the frame with probability $0.5$ and so the frame after measurement of a full generating set looks like $\mathcal{F}=[\bm{x}_s,\bm{z}]$ where $\bm{x}_s = H_X^T\bm{x}$ for a random binary vector $\bm{x}$. As mentioned previously, relative to a given $n$-qubit code any length-$n$ $\bm{z}$ can be decomposed into a combination of logicals, stabilizers and errors, but in this case the term ``error'' is misleading as $\bm{z}$ typically anticommutes with stabilizers even in the noiseless case. Instead we refer to this ``error'' term as a \textit{byproduct operator}~\cite{gaugingmeasurement} and write the decomposition of $\bm{z}$ as $\bm{z} = \bm{z}_l + \bm{z}_s + \bm{z}_b$.
\begin{center}
\begin{tcolorbox}[
  colback=white!95!black,
  colframe=white!20!black, 
  boxrule=0.5pt,
  width=0.7\linewidth,
  arc=2mm,
  boxsep=0pt,
  halign=center
]
    A choice of byproduct operator $\bm{z}_b$ fixes the logical class of $\bm{z}_l$ and we can always choose $\bm{z}_b$ such that $\bm{z}_l$ is trivial.
\end{tcolorbox}
\end{center}
The ability to choose a correction for $\bm{z}_b$ that trivialises $\bm{z}_l$ will be central to our ability to declare success or failure during decoding, and will be discussed in detail later.
For now, though, we assume such a correction has been computed and applied and so the Pauli frame is reduced to $\mathcal{F} = [\bm{x}_s,\bm{z}_s]$.

In the case of non-zero noise some set of random Pauli errors occurs in the circuit, and the error on the final state can be calculated by propagating these errors forward to the circuit's end. Because measurement of Pauli operators only requires Clifford gates this final error will also be Pauli, and is computed naturally by the previously defined action of Pauli errors and Clifford gates on the Pauli frame. We then end with a frame $\mathcal{F} = [\bm{x}_s + \bm{x}_e, \bm{z}_s + \bm{z}_e]$, and as errors that differ only by stabilizers are equivalent we can combine the stabilizer and error terms to get an equivalent frame $\mathcal{F} = [\bm{x}_e',\bm{z}_e']$. Moving forward, we can therefore consider frames that contain only error components.

For fault-tolerant initialisations of other Pauli eigenstates the procedure works in the same way.
After initialisation the frame might contain a non-trivial logical component which can be trivialised with an appropriate correction.

\subsubhead{Can we learn all measurement outcomes?}\\
The stabilizer measurements that prepare the state return a set of flips which in theory should then be compared to some set of reference outcomes as is done in \texttt{stim}. However, as we know that the true noiseless syndrome in this case should be fully random we can simply choose all reference outcomes to be $+1$ and take the flips as our syndrome directly.

\subsubsection*{Identity, Pauli and Clifford gates}
For now we specifically consider unitary implementations of these operations. Non-unitary implementations, e.g. surgery~\cite{CKBB, gaugingmeasurement, homological-measurement, Cowtan2026parallel, fastsurgeryIceberg, cowtanFast}, will be covered when we consider logical measurements. 

\subsubhead{What happens to the frame?}\\
Thanks to our choice of byproduct correction during state preparation we can assume a frame $\mathcal{F} = [\bm{x}_e, \bm{z}_e]$ as input to the operation. Logical identity blocks, i.e. rounds of stabilizer or Floquet measurement, multiply new stabilizers into the frame but do not change the logical content.
Logical Pauli gates do not affect the frame and so do not create a non-trivial logical component.
Logical Clifford operations do affect the frame, but always map stabilizers to stabilizers, logicals to logicals, and errors to errors. Hence, they cannot create a non-trivial logical component as logical identity is always mapped to logical identity.
The frame at the end of the block then once again contains only error components which depend only on the error component at the start and on any additional errors which happen throughout, all of which will be correctly tracked within the circuit as all physical operations are Clifford. 

\subsubhead{Can we learn all measurement outcomes?}\\
Tracking within the block is guaranteed to be correct and is identical to what is already done in \texttt{stim}, so provided the error component of the frame was tracked correctly prior to the start of the operation we can correctly infer all stabilizer measurement outcomes during or immediately following the block. If we know that these should all be $+1$ in the noiseless case, e.g. with repeated measurements of stabilizers on a perfect codestate, then we can once again interpret the flips directly as measurement outcomes. In more general dynamical protocols there may also be measurement outcomes that are either fully random or deterministically $-1$~\cite{delfosse2023spacetime}.
However, provided these do not perform logical measurements their outcomes will not be correlated with the input state, and a set of noiseless reference outcomes can be obtained by simulating the Clifford block using \texttt{stim} and a stabilizer input state. 

\subsubsection*{Non-Clifford gates}

We consider third-order blocks that implement a diagonal gate in the third level of the Clifford hierarchy on CSS codes.
This includes both unitary and non-unitary logical actions.
Ref.~\cite{companion-nonclifford} provides a detailed treatment of how to estimate the failure rates of these blocks reliably, independently of the logical state.
Upon success, the method returns a logically trivial Pauli frame with which we can interface the simulation with other logical blocks.
Here, we give a high-level description of how the transformation on the frame is obtained in the method described in Ref.~\cite{companion-nonclifford}.

\subsubhead{What happens to the frame?}\\
Before feeding the Pauli frame into the non-Clifford simulator, we must compute and add to the frame a correction $\bm{x}_c$ for $X$ errors that occurred earlier in the circuit (see Sec.~\ref{subsection:decoding}).
This corrected frame is then given as input to the third-order block.
In the general case, it proceeds by reliably sampling the error and measurement distribution within the block, and computing an $X$-like correction for the bulk of the block.
If this correction leads to a total $X$-error configuration that fulfils condition C~\cite{companion-nonclifford}, the correction was successful and the simulator continues to sample a $Z$-error configuration faithfully, including twisted errors.
Both $X$-like and $Z$-like errors can be propagated to the output boundary of the third-order block, yielding the updated output frame.
Since the sampling of the $Z$-like errors contains random twisted errors the frame transformation is non-deterministic, in stark contrast to Clifford blocks.

\subsubhead{Can we learn all measurement outcomes?}\\
In the event that the block returns a Pauli frame rather than reporting a logical failure our method from Ref.~\cite{companion-nonclifford} also returns a list of outcomes for the measurements that occurred inside the block.
The sampled outcomes follow the correct distribution, as long as the measurements do not contain any information about the logical state, or action of the block.
If $X$-decoding succeeds this only happens for third-order blocks that implement logical Clifford measurements, as the logical outcome can be inferred from the physical measurement outcomes.
In practice, such blocks are used when additional information about the logical outcome distribution is known, similar to logical Pauli measurements.
We discuss their treatment in the next section.

\subsubsection*{Logical measurements}
In modern fault-tolerant architectures logical measurements play an important role, even for a unitary computation, since most logical gates are performed in a Pauli-based computation scheme~\cite{bravyi2016trading}.
Here, we discuss how to treat logical blocks that implement physical Pauli measurements.

\subsubhead{What happens to the frame?}\\
Assume once again an input frame $\mathcal{F} = [\bm{x}_e,\bm{z}_e]$. The exact effect on the frame depends on the details of how the measurement is implemented. An idealised direct logical measurement would probabilistically add the measured logical into the Pauli frame as described in \cref{subsection:physical_operations}, but such measurements are usually not fault tolerant.
Instead, logical measurements are performed either by transversal measurements of a full codeblock or via surgery techniques.
In the former each measured single-qubit operator is probabilistically added to the frame, and in the latter case each measured stabilizer is probabilistically added to the frame.
This means we can write the output frame as either $[\bm{x}_e + \bm{x}, \bm{z}_e]$ or $[\bm{x}_e, \bm{z}_e + \bm{z}]$ for $X$ or $Z$ measurements respectively, where $\bm{x}$ or $\bm{z}$ describes a random product of the measured operators.
These operators have decompositions $\bm{x} = \bm{x}_l + \bm{x}_b$ and $\bm{z} = \bm{z}_l + \bm{z}_b$ in terms of the logicals of the pre-measurement code and a byproduct operator.
This makes logical measurements the only non-error process after state preparation that can result in a non-trivial logical component of the Pauli frame.

For transversal measurements this is not an issue, as the logical operators that can be added are precisely those that have no effect on the logical measurement outcome, and as a transversal measurement represents the end of a codeblock's life within a circuit we do not need to worry about how this might affect future operations or measurements on this code. This is not the case for code surgery, where the codes involved will typically remain active in the circuit after the measurement. To remove these randomly added logicals we can use the same method as in state preparation; we choose a correction for the byproduct operator that trivialises $\bm{z}_l$. The process for this is exactly the same as in state preparation, and will be discussed in the next subsection as part of decoding.
This process removes information about logical operators in the frame prior to measurements.
These are, however, not important as they never correspond to logical errors in the circuit.

\subsubhead{Can we learn all measurement outcomes?}\\
The short answer is \emph{no}, as this would imply the ability to accurately learn logical measurement outcomes and thus to efficiently sample from the output of a universal quantum circuit. However, measurement outcomes that do not reveal logical information can still be learned accurately. 

For instance, in the case of a transversal $Z$ measurement all $Z$ stabilizer eigenvalues immediately prior to the measurement can be accurately reconstructed as products of the single-qubit measurement outcomes, as these depend only on $\bm{x}_e$ and we can once again assume all reference outcomes \textit{for stabilizers} to be $+1$. Notice that we cannot take all reference outcomes \textit{for the single-qubit measurements} to be $+1$ as in a real circuit these will depend on the logical state of the code prior to measurement. 

For a similar reason, we cannot in principle accurately learn all stabilizer measurements in the case that the logical measurement is performed via code surgery, as the noiseless stabilizer outcomes reveal information about the logical state and we cannot generally take all reference outcomes to be $+1$. However, in practice logical measurements by code surgery are typically used as tools for implementing unitary quantum logic rather than for reading out quantum information and so should not reveal information about the encoded state.
In these cases, the distribution of logical outcomes should then be independent of this state.
As a result, we \textit{can} take all reference outcomes to be $+1$ in most cases of practical interest and can infer all stabilizer measurement outcomes from the Pauli frame as in the case of transversal measurement. Exceptions include logical readout in non-CSS codes, which do not always admit fault-tolerant transversal measurements and may require surgery for readout also. In such cases we cannot always correctly sample all stabilizer outcomes but can still sample all detector outcomes, as these depend only on correlations between repeated measurements of these stabilizers and not on their initial values.

\subsection{Decoding}
\label{subsection:decoding}

We have seen that we can accurately sample stabilizer outcomes and Pauli error configurations in circuits built from a computationally universal set of logical blocks, and now need to consider the question of how this information can be used in decoding, and how the success or failure of this decoding can be decided. We begin with a review of the standard approach to decoding Clifford circuits, then examine how decoding is altered by the introduction of non-Clifford blocks, and finally discuss the problem of decoding the byproduct operator produced during state initialisation and code switching, e.g. surgery, and how this allows us to ultimately infer decoding success or failure.

\subsubsection*{Decoding Clifford circuits}
Given a history of syndrome measurement outcomes for a fault-tolerant Clifford circuit the standard approach to performing decoding is to construct a \textit{detector-error model} (DEM~\cite{gidney_stim_2021}), which is a bipartite graph containing two kinds of vertices: \textit{detectors}, which correspond to sets of measurements for which the joint outcome is deterministic in the absence of physical errors, and \textit{errors}, which correspond to physical errors that can alter the outcomes of detectors. If an error alters the outcome of a detector then they are connected in the DEM by an edge. Detectors for which the outcome has been altered are said to be ``flipped'', and the set of detectors flipped by an error is called the \textit{syndrome}. A \textit{decoder} is an algorithm for identifying a subset of errors with a syndrome that matches the observed syndrome. In Pauli frame simulations of these circuits there is also a third type of object in the DEM, called an \textit{observable}. Observables can be thought of as special detectors which are not used in decoding, and are instead used to assess decoding success or failure; the decoder calculates a correction using only the flips of the detectors, but decoding only succeeds if this correction also matches the flips of the observables. 

In a real fault-tolerant circuit observables typically correspond to the outcomes of logical measurements. When a measurement is made, either transversally or via some form of surgery, we can compute a reliable logical outcome by taking the unreliable outcome obtained from a product of physical outcomes and then flipping this if our decoder's correction flipped the observable. In a large computation containing many logical measurements we can often simplify decoding by considering observables independently and decoding each one using a subgraph of the DEM containing only the errors that can flip that particular observable. Note that the sets of errors in these subgraphs do not all need to be disjoint, and care must be taken to ensure that different decoders do not disagree on corrections in shared regions of the DEM~\cite{gidney_stim_2021}. Note also that it is not necessary to actually compute a physical correction or apply it to the device. Knowing how the correction computed from the DEM flips the observable is sufficient. 

\subsubsection*{Decoding universal circuits}
In Clifford circuits under stochastic Pauli noise every error has a deterministic effect on the measurements, as Pauli errors remain Pauli as they propagate through the circuit.
This is no longer the case in circuits containing non-Clifford gates.
In particular, the combined action of propagation of a Pauli through a non-Clifford followed by projective measurements can implement a non-deterministic Pauli-to-Pauli map.
In more general third-order circuits that measure non-Pauli stabilizers to implement non-trivial logical operations, similar error-propagation behaviour can be derived~\cite{companion-nonclifford}. Despite this, the measurement outcomes are still subject to (affine) linear constraints leading to a well-defined decoding problem on detectors, similar to the Clifford case. However, the non-deterministic behaviour of Pauli errors in third-order circuits has implications both for when and how decoding must be performed, and also how success or failure is declared. 

\subsubhead{Physical corrections}\\
Firstly, unlike in Clifford circuits, it now becomes necessary to actually apply a physical correction to the device.
This is required even in the noiseless case, since e.g. preparing the state $\ket{\overline{+}}$ results in a byproduct operator $X[\bm{x}_b]$ with typical weight $n/2$.
If this state were given directly as input to a non-Clifford block, the state would be affected by a \emph{non-deterministic} error after measuring the $X$ stabilizers.
For a transversal logical gate this error would have typical weight $n/4$ on the output state.
This error cannot be reliably decoded and so represents a logical failure.
In other words, the non-Clifford block is only a fault-tolerant operation when the $X$ error on the input is sparse.
Before applying the block we must therefore perform a two-step decoding:
\begin{itemize}
    \item decode using the DEM and identify a correction in the circuit consistent with the observed syndrome
    \item propagate these corrections forwards to the input boundary of the non-Clifford block, and physically apply all resulting $X$ operators to the qubits.
\end{itemize}
It is not necessary to consider the entire history of $X$ corrections in the DEM for the propagation step.
Since $X$ corrections are physically applied at the start and within non-Clifford blocks only the $X$ corrections between consecutive non-Clifford blocks need to be propagated.
Additionally, for a given circuit it is possible to precompute a decoding subgraph containing only the faults which produce $X$ errors at the input boundary, and decode only on this subgraph. 

We must also make some modifications to the $X$-error decoding problem in the region of the non-Clifford
block by introducing closed boundary conditions at both the input and output.
These boundaries are defined by the removal from the DEM of all faults that lie within the block.
At the input boundary, this ensures that the physical correction applied before the block returns us to the codespace of the input code.
At the output boundary, corrections computed for the interior of the block are used to infer outcomes for measurements that are also part of the decoding problem outside the block. 
The $X$-error decoding within the block is performed by another method, see Ref.~\cite{companion-nonclifford}, so $Z$ detectors fully within the block are not part of the input to the decoder described here.

Given perfect decoding of the $X$-errors on the input and within the block the $Z$ decoding problem would be identical to the case for the circuit with all non-Clifford gates removed. 
In reality, there will be a residual $X$ error configuration which causes additional random $Z$ errors in the circuit, which we refer to as twisted errors in Refs.~\cite{highthresholdJIT, companion-nonclifford}.
The twisted errors can be captured exactly by a \emph{twisted DEM}, obtained by adding weight-0 faults to the DEM for $Z$ errors of the underlying CSS circuit.\footnote{Formally, the twisted DEM is related to the DEM of the underlying CSS circuit by a hypergraph morphism encoded in a matrix $K^b$ that is defined for any residual $X$-error configuration $b$ that fulfills condition C. For more details, we refer to Ref.~\cite{companion-nonclifford}.}
The decoder, however, only has access to partial information about the residual error. 
As a consequence, the decoder cannot operate on the true twisted DEM.
In principle, the third-order block is still fault tolerant if the $Z$ errors are decoded on an unmodified decoding graph~\cite{Scruby2022nonPauli, Beverland_2021}, albeit with reduced performance against $Z$ errors.
Ref.~\cite{highthresholdJIT} showed that the practical performance can be significantly improved using relatively simple heuristics based on the algebraic structure of the twisted errors.
We expect that further improvements can be made using the precise algebraic formulation of the twisted errors from Ref.~\cite{companion-nonclifford}.

We want to conclude this section by emphasising that the requirement to treat the decoding problem in this way has nothing to do with our simulation approach. It is a fundamental property of non-Clifford circuits which would be important when decoding universal logical circuits in practice on a real device.

\subsubhead{Logical outcomes}\\
Within a universal circuit logical outcomes are not usually deterministic in the absence of noise, in contrast to small toy experiments such as memory or stability experiments.
Hence, the standard procedure, based on identifying deterministic logical outcomes is not applicable to more general circuits.

The solution is to once again recall that the Pauli frame does not tell us anything about the real outcome of a measurement, only about whether that measurement was flipped relative to some noiseless reference. In particular, after computing and applying a correction, \textit{logical} measurements are flipped only if the Pauli frame contains a non-trivial logical operator that anticommutes with the measurement. We have seen in the previous subsection that, apart from errors, the only processes in the simulation which can map a logically trivial Pauli frame to a logically non-trivial one are state preparations and measurements. However, we have argued that in both cases a suitable choice of byproduct correction allows one to always remove the added logicals from the frame, resulting in a frame that is logically trivial. This then ensures that in a noiseless Pauli frame simulation \textit{every} logical measurement reports no flip. This in turn means that any flip we observe in the noisy circuit must be due to a decoding failure. In this way we do not require any knowledge of the real circuit outcome statistics in order to distinguish between decoding success and failure. All that remains is an explanation of how one can always choose the byproduct correction to achieve this. 
\newline

\subsubhead{Decoding the byproduct operator}\\
 In general there will be one byproduct correction for each prepared codeblock, and also for each logical measurement performed via surgery.
 All of these can be treated independently via the same method so for concreteness we consider the case of decoding a single byproduct operator occurring during $\ket{\overline{0}}^{\otimes k}$ state preparation. 

Immediately after preparing the physical qubits in $\ket{\overline{0}}^{\otimes n}$ and measuring a generating set of $X$ stabilizers the Pauli frame looks like
\begin{equation}
    \label{eq:frame_copy}
        \mathcal{F} = [..., \bm{z}_l + \bm{z}_s + \bm{z}_b + \bm{z}_e], 
\end{equation}
where we omitted the $X$ part of the frame as it is not relevant to the discussion.
From this point we will also omit $\bm{z}_s$ as it can be absorbed into any of the other terms without loss of generality.
At this point, the simulation saves a copy $\mathcal{F}_\textrm{copy}$ of the frame.
At some later point in the circuit the decoder will decide on a correction for the measurement errors that occurred in this first measurement round and these can be combined with the recorded measurement outcomes to obtain a corrected syndrome for $\bm{z}_b$. Note that this is a syndrome on the code-capacity decoding graph for the initialised code, not on the circuit's DEM since the byproduct operator occurs even in the noiseless circuit.
This decoding problem can then be solved to find a correction $\bm{z}_{b,\textrm{cor}}$ and while generally this correction will not be exact -- in the sense that $\bm{z}_b + \bm{z}_{b,\textrm{cor}}$ does not usually have trivial syndrome -- the fault tolerance of the circuit guarantees that a state $Z[\bm{z}_b]Z[\bm{z}_{b,\textrm{cor}}]\ket{\overline{\psi}}$ is far from the codespace only if a high-weight fault occurred.

Once we have chosen $\bm{z}_{b,\textrm{cor}}$, the next step is to correct the residual error in the noisy frame.
We first add $\bm{z}_{b,\textrm{cor}}$ into our stored copy of the noisy frame after initialisation, which becomes $\mathcal{F}_\textrm{copy} = [..., \bm{z}_l + \bm{z}_r]$ where we have packaged $\bm{z}_b + \bm{z}_e + \bm{z}_{b,\textrm{cor}}$ into a residual error $\bm{z}_r$. From this frame we compute a syndrome for $\bm{z}_r$, decode it and apply a correction $\bm{z}_{r,\textrm{cor}}$ to obtain $\mathcal{F}_\textrm{copy} = [..., \bm{z}_L]$ where $\bm{z}_L = \bm{z}_l + \bm{z}_r + \bm{z}_{r,\textrm{cor}}$ is a logical operator. We now have a ``clean'' version of the noisy frame with a well-defined logical $Z$ class. 

The last step is to trivialise this frame, which can be done given access to a symplectically paired basis of logical $X$ and $Z$ operators for the prepared code.
To that end, we first check which logical $X$ operators anticommute with $Z[\bm{z}_L]$ and then apply their logical $Z$ counterparts, whose product we can call $Z[\bm{z}_L']$.
Our final choice of byproduct correction is then $\bm{z}_{b,\textrm{cor}} + \bm{z}_L'$. 

To see why this works let's return to the frame from \cref{eq:frame_copy} and imagine applying our chosen byproduct correction.
This yields a corrected Pauli frame
\begin{equation}
    \begin{split}
        \mathcal{F}_n &= [..., \bm{z}_l + \bm{z}_b + \bm{z}_e + \bm{z}_{b,\textrm{cor}} + \bm{z}_L'] \\
        &= [..., \bm{z}_l + \bm{z}_r + \bm{z}_L'] \\
        &= [..., \bm{z}_L + \bm{z}_{r,\textrm{cor}} + \bm{z}_L'] \\
        &= [..., \bm{z}_{r,\textrm{cor}}]
    \end{split}
\end{equation}
which only contains the error configuration $\bm{z}_{r,\textrm{cor}}$.
It's worth noting that $\bm{z}_{r,\textrm{cor}}$ is not guaranteed to belong to the same logical coset as $\bm{z}_r$. If many errors occur in the state preparation circuit then $\bm{z}_r$ can have high weight, and the code capacity decoder can fail to return a $\bm{z}_{r,\textrm{cor}}$ in the same logical coset. This is not an issue, as we are preparing logical $\ket{\overline{0}}^{\otimes k}$ and $\bm{z}_r$ and $\bm{z}_{r,\textrm{cor}}$ must be equivalent up to composition with some logical $Z$ operator, which will be a stabilizer of this state. 

Once this byproduct correction has been calculated it can be stored by the simulation until it is needed.
For example, if we wish to prepare a $\ket{\overline{+}}$ state and then immediately use it as input to a non-Clifford block the byproduct $X$ correction is needed straight away, as it is part of the calculation of the physical $X$ correction that must be applied to the qubits before applying any non-Clifford gates.
On the other hand, in a memory experiment where we prepare a state in $\ket{\overline{0}}$ and only care about the outcome of the logical $Z$ observable after some number of measurement rounds, the byproduct correction is never needed.

\subsection{Adaptive quantum circuits}
In this section we explain how the methods above can be used to simulate certain fault-tolerant circuits that include adaptive corrections based on logical measurement outcomes.
These types of operations feature prominently in state-injection and Pauli-product $T$-rotation gadgets.
To that end, we recall our original goal, formulated in Sec.~\ref{section:overview}.
To estimate the failure rates of entire logical circuits it suffices to sample random outcomes that determine the adaptive corrections at the start and pass the resulting logical circuit, including the resulting corrections, as a fixed input to a simulator that is composed of the components described in Sec.~\ref{section:blueprint} so far.

To see how this works, we consider a quantum circuit that acts unitarily on the encoded information. In this case, any mid-circuit measurement outcomes must be independent of the logical state being processed by the circuit, as they would otherwise lead to a non-unitary computation.
Assuming that we can sample from the outcome distribution for these measurements we can then decide on a set of outcomes ahead of time and pass the resulting circuit, with operations conditioned on the sampled outcomes, to our simulator. This is not a very harsh assumption, as the measurement outcomes in almost all commonly used adaptive quantum operations (state injection, teleportation, etc.) are fully random and independent and identically distributed (IID) and the sampling requires only a fair coin toss for each measurement. Even in non-uniform cases such as measurement of Clifford logicals~\cite{tiedinknots,bauer2025planar,zhu2026nonabelianqldpc,williamson2026fastmagic, newmagic} the distribution of measurement outcomes is often known. It is valid to choose these outcomes before running the simulator because, once again, we are free to choose a set of reference outcomes from the distribution of possible outcomes for these measurements how we like. Choosing which adaptive operations to apply ahead of time can be understood as choosing our reference outcomes at runtime so that reference outcome plus computed flip from the Pauli frame always gives the pre-chosen outcome. Note that this is on a purely conceptual level, and computation of these reference outcomes is not required in the actual simulation.

We propose drawing a random pattern for these conditioning measurements -- rather than choosing them all to be $+1$ as in previous cases, and so applying no adaptive operations -- as in most hardware platforms error rates from gates are higher than from idling, and so we do not expect the all $+1$s instance to produce logical error rates representative of a typical circuit instance. For maximum accuracy one should draw a new random circuit instance for each simulation shot, but the extent to which this is necessary in reality will likely depend on the specific circuit and noise model being studied. In \texttt{stim}, for example, it is significantly cheaper to compile the sampler for a circuit once and sample from it many times than to compile a new sampler for each shot, so it would be ideal if in practice only a few random circuit instances are required to approximate the failure rate of the general protocol accurately, but we leave this as a problem for future study. 

\section{Discussion}
\label{section:discussion}
We have presented a design for an efficient end-to-end simulator of error propagation and decoding in universal quantum circuits under stochastic Pauli noise.
We focus specifically on logical circuits built from Clifford operations and diagonal gates in the third level of the Clifford hierarchy, which form the basis of most modern fault-tolerant architectures.
To the best of our knowledge this is the first scalable  and efficient approach to the simulation of the performance of large-scale fault-tolerant algorithms.

While this paper presents more of a conceptual blueprint, the required operations are relatively simple and involve comparatively minor modifications to widely used Pauli-frame tracking methods.
Already today this would allow for an efficient estimation of failure rates of larger subroutines of logical circuits.
For example, it would be interesting to investigate the performance of an entire logical adder, or a quantum Fourier transform, implemented fault tolerantly.
This would likely require development of a streaming version of our simulator in which the classical descriptions of the physical circuits and detector error models are loaded dynamically to prevent infeasible memory usage.

For very large circuits, it may be necessary to develop  coarse-graining methods, where mid-sized subroutines are simulated on their own and fed to a larger-scale simulation module. Simulation methods of this kind have already been used for simulating large concatenated codes, where inner codes are simulated at the circuit level and coarse-grained for use in outer code simulations~\cite{yoked}.

Our simulator covers a large and important set of physical operations and logical blocks.
Non-Clifford blocks, however, remain restricted to diagonal non-Clifford gates.
Extending the scope of the non-Clifford subroutine would allow for an even larger set of operations, including a recently introduced scheme to measure non-diagonal Clifford logicals fault tolerantly~\cite{newmagic}.
It would also be interesting to explore the connection and compatibility with recently introduced tableau-based simulators for non-Clifford error-correcting circuits~\cite{yugopaper, cercelescu2026polynomialtimesimulationnoncliffordquantum}.

On a fundamental level, the working principle of our simulator highlights that physical quantities, such as measurement outcomes, are highly restricted in fault-tolerant circuits.
A more thorough understanding of information-theoretic properties of the measurement distribution in fault-tolerant circuits could provide insights into the study of highly entangled quantum dynamics beyond simple Clifford circuits, and could potentially be useful for the characterisation and simulation of more realistic noise in general error-correcting circuits.

\section*{Acknowledgements}
We thank the authors of Ref.~\cite{yugopaper} for useful discussions about their work. Parts of this work were completed at the Fault Tolerant Quantum Technologies workshop at the Centro de Ciencias de Benasque Pedro Pascual and the Fault Tolerant Logical Processing workshop at the Yukawa Institute for Theoretical Physics. 
JM acknowledges funding from the DFG (CRC183) and from Defence Science and Technologies Group (DSTG) and Advanced Strategic Capabilities Accelerator (ASCA) through its Emerging and Disruptive Technologies (EDT) Program.
This work was supported in part by Japan Science and Technology Agency (JST) as part of Adopting Sustainable Partnerships for Innovative Research Ecosystem (ASPIRE), Grant Number JPMJAP25A3.
All results presented in this paper were obtained without the help of any large language model.

\begin{appendix}
\section{Sampling twisted errors from the Pauli frame}
\label{app:ccz_sampling}

In Ref.~\cite{companion-nonclifford} we describe a method for sampling \textit{twisted errors} -- the non-deterministic $Z$ errors resulting from projection of Clifford errors -- in third-order circuits.
This method is general and efficient, but requires several steps, including Gaussian elimination on a sparse matrix defined on the whole non-Clifford block.
While further structure of the circuit, such as translational invariance, can be leveraged to improve on that worst-case scaling~\cite{highthresholdJIT}, a faster method for the generic case would be very beneficial.
In this appendix we present a method with which twisted errors can be sampled much more efficiently provided the third-order circuit satisfies an additional property.
We will show that this property is fulfilled in circuits that act with CCZ gates between disjoint codeblocks whose stabilizers are measured directly before and after applying the CCZs.
This method relies solely on the Pauli frame and samples twisted errors via simple gate-local Pauli frame updates of the same type used in \texttt{stim} for Clifford circuits.
Specifically, these rules are given by 
\[
    U_\textrm{NC}: [\bm{x},\bm{z}] \rightarrow [\bm{x},\bm{z}'], \textrm{ for } U_\textrm{NC}(\phi X[\bm{x}]Z[\bm{z}])U_\textrm{NC}^\dagger = \phi' X[\bm{x}]Z[\bm{z}']U_\textrm{C}'
\]
which uses the fact that third-level non-Clifford gates map Pauli operators to Clifford operators. In general we have $U_\textrm{NC}(\phi X[\bm{x}]Z[\bm{z}])U_\textrm{NC}^\dagger = \phi' U_\textrm{C}$ for some Clifford operator $U_\textrm{C}$. This can always be written as a product $\phi\prod_i U_P^i U_\textrm{C}^i$ where $U_P^i$ are Pauli operators and $U_\textrm{C}^i$ are non-Pauli Clifford operators. Then, because Clifford operators map Paulis to Paulis we can commute all the Pauli terms to the left and rewrite this as $\phi'\prod_i {U_P^i}' \prod_j U_\textrm{C}^j = \phi'X[\bm{x}']Z[\bm{z}']U_\textrm{C}'$. Finally we observe that $\bm{x} = \bm{x}'$ because $U_\textrm{NC}$ is diagonal.

For the rest of this section we consider a CSS stabilizer code $\mathcal{C}$ that admits a constant-depth unitary third-order logical circuit $\overline{U}_{NC}$. We consider a circuit that implements a round of direct syndrome extraction for $\mathcal{C}$, applies $\overline{U}_{NC}$, and then performs another round of syndrome extraction.
As long as the effective Clifford error does not have overlap with any logical $Z$ operator, it gets completely projected onto random Pauli errors, which we can capture in the Pauli frame.
We prove that in this case a simple update rule on the frame directly captures the correct statistics of the circuit, provided that the stabilizers are not flipped directly via the Clifford error.

\begin{theorem}
    \label{theorem:main}
    Consider a code $\mathcal{C}$ and unitary circuit $\overline{U}_{NC}$ composed of diagonal gates in the third level of the Clifford hierarchy and let $\ket{\tilde{\psi}} = X[\bm{x}_e]Z[\bm{z}_e]\ket{\overline{\psi}}$ be a noisy codestate of $\mathcal{C}$.
    Let $U_\textrm{C}$ be the Clifford operator defined by $\overline{U}_\textrm{NC}X[\bm{x}_e]Z[\bm{z}_e]\overline{U}_\textrm{NC}^\dagger = \phi_e X[\bm{x}_e]Z[\bm{z}_e']U_\textrm{C}$ with some global phase $\phi_e$, and $\mathcal{F}$ be the Pauli frame in the simulation immediately prior to application of the non-Clifford gate.
    If $U_\textrm{C}$ is supported on qubits that do not support any $Z$ logical of $\mathcal{C}$ then the measurement of a stabilizer $X[\hat{\bm{x}}_s]Z[\hat{\bm{z}}_s]$ of $\mathcal{C}$ on a copy of $\overline{U}_\textrm{NC}\ket{\tilde{\psi}}$ samples from the same outcome distribution as a flip of $X[\hat{\bm{x}}_s]Z[\hat{\bm{z}}_s]$ computed from $\overline{U}_\textrm{NC}(\mathcal{F})$, except in cases where $U_\textrm{C}^\dagger X[\hat{\bm{x}}_s]Z[\hat{\bm{z}}_s] U_\textrm{C}\ket{\overline{\psi}} = -\ket{\overline{\psi}}$.
\end{theorem}

\begin{proof}

    As $\mathcal{C}$ is CSS it suffices to prove the statement for $X$ and $Z$ stabilizers separately. The proof for $Z$ is trivial as $\overline{U}_\textrm{NC}$ is diagonal and commutes with all $Z$ stabilizers, so we only need to consider $X$ stabilizers. First, 
    \begin{equation}
        \overline{U}_\textrm{NC}\ket{\tilde{\psi}} 
        = \phi_e X[\bm{x}_e]Z[\bm{z}_e']U_\textrm{C}\overline{U}_\textrm{NC}\ket{\overline{\psi}}
        = \phi_e X[\bm{x}_e]Z[\bm{z}_e']U_\textrm{C}\ket{\overline{\psi'}}
    \end{equation}
    as $\overline{U}_\textrm{NC}$ is a logical operation of $\mathcal{C}$ and so preserves the codespace. 
    Because $U_\textrm{C}$ is Clifford we can think of it as a transformation of the stabilizer group of $\mathcal{C}$ and then, by the standard stabilizer formalism~\cite{gottesman1997stabilizercodesquantumerror}, measurement of the original stabilizers must return either a deterministic measurement outcome or a completely random measurement outcome. To check which for a given stabilizer $X[\hat{\bm{x}}_s]$ we can apply this stabilizer to the state (neglecting the phase $\phi_e$) to get
    \begin{equation}
        X[\hat{\bm{x}}_s]X[\bm{x}_e]Z[\bm{z}'_e]U_\textrm{C}\ket{\overline{\psi}} = (-1)^{\hat{\bm{x}}_s \cdot \bm{z}'_e} X[\bm{x}_e]Z[\bm{z}'_e]U_\textrm{C} (\phi X[\hat{\bm{x}}_s]Z[\bm{z}])\ket{\overline{\psi}}
    \end{equation}
    where $\phi X[\hat{\bm{x}}_s]Z[\bm{z}] = U_\textrm{C}^\dagger X[\hat{\bm{x}}_s]U_\textrm{C}$. Notice that $\phi Z[\bm{z}]$ cannot be a logical operator because it can be created only within the support of $U_\textrm{C}$, which is cleanable. Thus we have three possible cases:

    \begin{enumerate}
        \item If $\phi Z[\bm{z}]$ is a stabilizer of $\mathcal{C}$ then $X[\hat{\bm{x}}_s]$ measures $(-1)^{\hat{\bm{x}}_s \cdot \bm{z}'_e + \hat{\bm{z}}_s \cdot \bm{x}'_e}$
        \item If $-\phi Z[\bm{z}]$ is a stabilizer of $\mathcal{C}$ then $X[\hat{\bm{x}}_s]$ measures $-(-1)^{\hat{\bm{x}}_s \cdot \bm{z}'_e + \hat{\bm{z}}_s \cdot \bm{x}'_e}$
        \item Otherwise $X[\hat{\bm{x}}_s]$ measures a random outcome.
    \end{enumerate}

    Now consider the effect on the Pauli frame $\mathcal{F}$. After the stabilizer measurement round prior to application of $\overline{U}_\textrm{NC}$ we have $\mathcal{F} = [\bm{x}_e + \bm{x}_s, \bm{z}_e + \bm{z}_s]$ where $\bm{x}_s$ and $\bm{z}_s$ are random elements of the stabilizer group of $\mathcal{C}$. Then $\overline{U}_\textrm{NC}$ transforms the frame as $\overline{U}_\textrm{NC}(\mathcal{F}) = [\bm{x}_e + \bm{x}_s, \bm{z}_e' + \bm{z}_s'']$ defined by
    \begin{equation}
        \begin{split}
        \overline{U}_\textrm{NC}X[\bm{x}_e + \bm{x}_s]Z[\bm{z}_e + \bm{z}_s]\overline{U}_\textrm{NC}^\dagger 
        = ~&\overline{U}_\textrm{NC}X[\bm{x}_e]Z[\bm{z}_e]X[\bm{x}_s]Z[\bm{z}_s]\overline{U}_\textrm{NC}^\dagger \\
        \propto~ & X[\bm{x}_e]Z[\bm{z}_e']U_\textrm{C}\overline{U}_\textrm{NC}X[\bm{x}_s]Z[\bm{z}_s]\overline{U}_\textrm{NC}^\dagger \\
        \propto~ & X[\bm{x}_e]Z[\bm{z}_e']U_\textrm{C}X[\bm{x}_s]Z[\bm{z}_s']U_\textrm{Cs} \\
        \propto~ & X[\bm{x}_e]Z[\bm{z}_e']X[\bm{x}_s]Z[\bm{z}_s'']U_\textrm{C}U_\textrm{Cs},
        \end{split}
    \end{equation} 
    where $\bm{z}_s'$ is a $Z$ stabilizer and $U_\textrm{Cs}$ is a Clifford stabilizer (since $\overline{U}_\textrm{NC}$ preserves the codespace). The flip of $X[\hat{\bm{x}}_s]$ computed from $\overline{U}_\textrm{NC}(\mathcal{F})$ is $(-1)^{\hat{\bm{x}}_s \cdot (\bm{z}_e' + \bm{z}_s'')} = (-1)^{\hat{\bm{x}_s} \cdot \bm{z}_e'}(-1)^{\hat{\bm{x}_s} \cdot \bm{z}_s''}$ where $\hat{\bm{x}}_s \cdot \bm{z}_s''$ can be calculated as the group commutator\footnote{The group commutator is defined as $[A,B] = ABA^{-1}B^{-1}$}
    \begin{equation}
        \begin{split}
        [X[\hat{\bm{x}}_s], Z[\bm{z}_s'']] &= [X[\hat{\bm{x}}_s], X[\bm{x}_s]Z[\bm{z}_s'']] \\
        &= [X[\hat{\bm{x}}_s], U_\textrm{C}X[\bm{x}_s]Z[\bm{z}_s']U_\textrm{C}^\dagger]
        \end{split}
    \end{equation}
    Because any group commutator of Paulis equals $\pm I$ we can equivalently calculate
    \begin{equation}
        \begin{split}
            U^\dagger [X[\hat{\bm{x}}_s], UX[\bm{x}_s]Z[\bm{z}_s']U^\dagger] U 
            &= U^\dagger (X[\hat{\bm{x}}_s] UX[\bm{x}_s]Z[\bm{z}_s']U^\dagger X[\hat{\bm{x}}_s] U Z[\bm{z}_s']X[\bm{x}_s]U^\dagger) U \\
            & = [U^\dagger X[\hat{\bm{x}}_s]U, X[\bm{x}_s]Z[\bm{z}_s']] \\
            & = [X[\hat{\bm{x}}_s]Z[\bm{z}], X[\bm{x}_s]Z[\bm{z}_s']] \\
            & = [Z[\bm{z}], X[\bm{x}_s]]
        \end{split}
    \end{equation}
    Recall that $X[\bm{x}_s]$ is a randomly selected $X$ stabilizer. In cases $1$ and $2$ as listed above $[Z[\bm{z}], X[\bm{x}_s]] = +I$ for any $X[\bm{x}_s]$ and our inferred measurement outcome is given simply by $(-1)^{\hat{\bm{x}}_s \cdot \bm{z}_e'}$. In case $3$ we have $[Z[\bm{z}], X[\bm{x}_s]] = \pm I$ randomly depending on whether $X[\bm{x}_s]$ commutes with $Z[\bm{z}]$ or not\footnote{Any Pauli operator that anticommutes with at least one stabilizer partitions the stabilizer group into a subgroup of all stabilizers that commute with it and a single coset of all stabilizers that do not. Thus, by choosing a random stabilizer and computing the commutator, we obtain a fully random outcome.}, and so our inferred measurement outcome is also random. Thus in cases $1$ and $3$ we infer the correct measurement outcome but in case $2$ we infer the incorrect outcome. 
\end{proof}

\begin{remark}
    The inclusion of random elements of the stabilizer group in the Pauli frame is not artificial; such elements are naturally added to the frame every time we measure the stabilizers of the code which must have taken place at least once before the application of the gate in order for us to be in the correct codespace. 
\end{remark}

\begin{remark}
    Case $2$ in the proof of Thm.~\ref{theorem:main} above corresponds to a subset of the twisted detectors where the affine shift in the twisted decoding problem as presented in~\cite{companion-nonclifford} is non-zero. Other twisted detectors on which the affine shift is non-zero correspond to stabilizers flipped by the $Z$ errors $Z[\bm{z}_e + \bm{z}_e']$. The errors $Z[\bm{z}_s'']$ are the sampled twisted errors not associated with the affine shift.
\end{remark}

Given this result, it is natural to ask in which settings case $2$ presents an issue, and in which cases it does not. One interesting class of circuits for which case $2$ is not an issue is any CCZ circuit where CCZ gates always act between qubits in different codeblocks, such as copy-cup CCZ gates~\cite{breuckmann2025cupsgatesicohomology, tiew2026copycupgatestensorproducts}.
This is quite a powerful result, as it allows one to avoid the need for Gaussian elimination as discussed in Ref.~\cite{companion-nonclifford} and accurately sample all twisted errors directly from the Pauli frame locally in these cases.

\begin{theorem}\label{thm:thm2}
    Let $\overline{U}_\textrm{CCZ}$ be a logical operation implemented by physical CCZ gates with each gate acting on qubits of three different codeblocks. Then, any $X$ error $X[\bm{x}_e]$ in these codes propagates to a Clifford error $U_{CZ}$ composed of CZ gates.
    None of these Clifford errors have the property that $U_\textrm{CZ}X[\bm{x}_s]U_\textrm{CZ} = -X[\bm{x}_s]Z[\bm{z}_s]$, where $X[\bm{x}_s]$ and $Z[\bm{z}_s]$ are stabilizers of $\mathcal{C}$.
\end{theorem}

\begin{proof}
    By the equivalence $\textrm{CCZ}(X\otimes I \otimes I)\textrm{CCZ} = X \otimes \textrm{CZ}$ we can see that any single-qubit $X$ error in one code results in one or more CZ errors acting between the other two codes after application of $\overline{U}_\textrm{CCZ}$. Assume there exists a product of such CZ errors, $U_\textrm{CZ}$, with the property $U_\textrm{CZ} X[\bm{x}_s] U_\textrm{CZ} = -X[\bm{x}_s]Z[\bm{z}_s]$ for some pair of $X$ and $Z$ stabilizers defined by $\bm{x}_s$ and $\bm{z}_s$. Without loss of generality, call the code containing the original $X$ error code $1$, and the codes containing the resulting CZ error codes $2$ and $3$. Now notice that 
\begin{equation}
    \begin{split}
        CZ(I \otimes I)CZ &= I \otimes I \\
        CZ(X \otimes I)CZ &= X \otimes Z \\
        CZ(I \otimes X)CZ &= Z \otimes X \\
        CZ(X \otimes X)CZ &= - XZ \otimes XZ \\
    \end{split}
\end{equation}
and thus a CZ gate can create a minus sign only when both qubits in the support of the gate are also in the support of an $X$ operator. Thus in order to have $U_\textrm{CZ}$ such that $U_\textrm{CZ} X[\bm{x}_s] U_\textrm{CZ} = -X[\bm{x}_s]Z[\bm{z}_s]$, $X[\bm{x}_s]$ must be a product of stabilizers from codes $2$ and $3$. Call these $X[\bm{x}_s^2]$ and $X[\bm{x}_s^3]$, with $\bm{x}_s = \bm{x}_s^2 + \bm{x}_s^3$, and let $m$ be the number of CZ gates in $U_\textrm{CZ}$ that have support on qubits of both $X[\bm{x}_s^2]$ and $X[\bm{x}_s^3]$. Notice also that $U_\textrm{CZ} X[\bm{x}_s^2]U_\textrm{CZ} = X[\bm{x}_s^2]Z[\bm{z}_s^3]$ and $U_\textrm{CZ} X[\bm{x}_s^3]U_\textrm{CZ} = X[\bm{x}_s^3]Z[\bm{z}_s^2]$ where $\bm{z}_s = \bm{z}_s^2 + \bm{z}_s^3$. To have an overall minus sign in $U_\textrm{CZ} X[\bm{x}_s] U_\textrm{CZ} = -X[\bm{x}_s]Z[\bm{z}_s]$ we must have that $m$ is odd. However, the size of the intersection between $X[\bm{x}_s^2]$ and $Z[\bm{z}_s^2]$ (or between $X[\bm{x}_s^3]$ and $Z[\bm{z}_s^3]$) has parity equal to the parity of $m$, and so this cannot be odd or the stabilizers of the individual codes would not commute. Thus $U_\textrm{CCZ}$ cannot create CZ errors of this type. 
\end{proof}
\end{appendix}

\bibliographystyle{quantum}
\bibliography{references}

\end{document}